\documentclass[twocolumn,showpacs,pra,aps,superscriptaddress]{revtex4}
\usepackage{ulem,bm}
\usepackage{amsmath, upgreek, amssymb, graphicx, paralist, gensymb,epstopdf}
\usepackage{array,epsfig,amsmath,amssymb}
\usepackage{amsmath,epstopdf}
\usepackage{graphicx}
\usepackage{dsfont}
\usepackage{color}

\def\bra#1{\langle #1|}
\def\ket#1{|#1 \rangle}

\begin{document}

\title{Testing genuine multipartite nonlocality in phase space}

\author{Seung-Woo Lee}
\affiliation{Center for Macroscopic Quantum Control, Department of
Physics and Astronomy, Seoul National University, Seoul, 151-742,
Korea}

\author{Mauro Paternostro}
\affiliation{Centre for Theoretical Atomic, Molecular and Optical
Physics, School of Mathematics and Physics, Queen's University
Belfast, BT7 1NN Belfast, United Kingdom} 

\author{Jinhyoung Lee}
\affiliation{Department of Physics, Hanyang University, Seoul
133-791, Korea}
\affiliation{Center for Macroscopic Quantum Control, Department of
Physics and Astronomy, Seoul National University, Seoul, 151-742,
Korea}

\author{Hyunseok Jeong}
\affiliation{Center for Macroscopic Quantum Control, Department of
Physics and Astronomy, Seoul National University, Seoul, 151-742,
Korea}

\date{\today}

\begin{abstract}
We demonstrate genuine three-mode nonlocality based on phase space formalism. A Svetlichny-type Bell inequality is formulated in terms of the $s$-parameterized quasiprobability function. We test such tool using exemplary forms of three-mode entangled states, identifying the ideal measurement settings required for each state. We thus verify the presence of genuine three-mode nonlocality that cannot be reproduced by local or nonlocal hidden variable models between any two out of three modes. In our results, GHZ- and W-type nonlocality can be fully discriminated. We also study the behavior of genuine tripartite nonlocality under the effects of detection inefficiency and dissipation induced by local thermal environments. Our formalism can be useful to test the sharing of genuine multipartite quantum correlations among the elements of some interesting physical settings, including arrays of trapped ions and intracavity ultracold atoms.  
\end{abstract}


\maketitle

Quantum nonlocality is one of the most fundamental features of quantum mechanics. It refers to the correlations that cannot be explained by local hidden-variable models that satisfy a set of constraints epitomised by so-called Bell inequalities~\cite{Bell64}. The violation of a Bell inequality reveals the existence of nonlocality in a given quantum mechanical state~\cite{CHSH69,CH74,Aspect1982}. 

While originally formulated for bipartite systems, Bell-like inequalities have been extended to the multipartite scenario, a noticeable example being embodied by the well-known inequality proposed by Mermin and Klyshko (MK)~\cite{merminkla}. However, the violation of a MK-type inequality by a multipartite state does not necessarily imply the existence of {\em genuine} multipartite nonlocality, as this test can be flasified by nonlocal correlations in any reduction of the system's components. In order to demonstrate genuine tripartite nonlocality, another type of Bell inequalities should be thus considered such as the one formulated by Svetlichny \cite{Svetlichny}, which rules out both local and nonlocal hidden variable model possibly imposed on any sub-parties. It was also noted that a stronger violation of the MK type can demonstrate genuine nonlocality for the cases with even number of parties \cite{Collins02}. Experimental demonstrations of genuine multipartite nonlocality were firstly achieved by strong violations of a MK inquality with four photon polarization entanglements \cite{Zhao03}. A violation of Svetlichny type inequality was experimentally demonstrated recently with GHZ-type photon polarization entangled states \cite{Lavoie09}. A generalised version of Svetlichny inequality was recently proposed and studied \cite{Bancal11}.

In this paper, motivated by the growing experimental capabilities of controlling and manipulating the state of tripartite quantum systems, in the optical lab~\cite{Braunstein05} and beyond, we address the formulation of genuine three-mode continuous variable (CV) nonlocality tests in phase space. While the phase space provides a natural arena for the description of the state of multi-mode CV systems, it also provides us with powerful tools for the analysis of the quantum correlation features within a given state 
and the possibility to investigate the quantum-to-classical transition in a transparent, intuitive way. While, historically, the completeness of quantum mechanics has been questioned by Einstein, Podolsky and Rosen using an argument related to position and momentum of a CV system~\cite{EPR35}, the first formulation of a Bell-like nonlocality test in phase space was provided by Banaszek and Wodkiewicz using the Wigner and Husimi $Q$-function~\cite{Banaszek99}. This approach has been generalized and extended in several directions, including  the provision of phase-space tests valid for high-dimensional systems~\cite{WSon06,SWLee09-2,SWLee11} and the addressing of general $s$-parameterized quasiprobability functions~\cite{Cahill69,SWLee09,SWLee10}. Although a considerable body of work has been produced to address issues of non locality in multi-mode CV states ~\cite{PvanL00,PvanL01,PvanL02,Ferraro05,Acin09,McKeown10,Li11}, a self-consistent formulation of genuine multimode nonlocality tests in phase space is still missing. 

In this paper, we fill this important gap by formalising MK and Svetlichny-type Bell inequalities in terms of the $s$-parameterized quasiprobability functions~\cite{Cahill69}, and thus opening the way to the experimental test of multipartite nonlocality of CV systems in their phase space. In order to assert the usefulness of our formal tools, we study in depth the tripartite nonlocal character of three paradigmatic states: a three-mode single-photon entangled state, a tripartite squeezed vacuum state, and the entangled coherent state of three bosonic modes. By exploiting the generality of our formulation and exploring the properties of these classes of states, we build up complete tripartite-nonlocality phase diagrams, identifying regimes where  the presence of genuine three-mode nonlocality is guaranteed and investigating its character in terms of sharing structure of quantum correlations. 
We complete our analysis by analysing the effects of both local damping and detection inefficiencies on the genuine tripartite nonlocal nature of a given state, proving that our inequalities are flexible enough to incorporate such detrimental effects in a natural way. Although we do not explicitly address it here, the extension of our formalism to the larger registers of modes is fully straightforward. Moreover, as we show in this paper, this investigation holds the promises to provide a useful tool for the characterization of the quantum correlation-sharing structure in physical settings of current enormous technological interests.  

The remainder of this paper is organized as follows. Bell-like inequalities for multipartite scenarios  are briefly reviewed in Sec.~\ref{section:Svet}. This is then followed by our formulation of phase-space tests based on generalized quasiprobability functions, as presented in Sec.~\ref{section:PSformalism}. In Sec.~\ref{section:genuinenonlocality} we pass to investigate the violation of such inequality by  the abovementioned classes of three-mode entangled states, whose tripartite nonlocal character is studied against relevant sources of detrimental effects in Sec.~\ref{section:noiseffcts}. Sec.~\ref{section:Settings} is then devoted to the identification and brief discussion of a series of physical setups where our study will be invaluable for the identification of the proper sharing structure of quantum correlations. Finally, in Sect.~\ref{section:discon} we draw our conclusions and highlight potential lines of developments.

\section{Genuine multipartite nonlocality}
\label{section:Svet}

In this Section, we discuss the formal approach to the derivation of  Bell inequalities for the test of
multipartite nonlocality. In order to fix the ideas, we consider explicitly the case of a tripartite system, although the 
discussion here is straightforwardly generalised to the more than three subsystems. 

It is instructive to briefly revise the two-party case, first. 
Let us thus assume that two particles, whose state can be repeatedly and identically prepared, are distributed to two remote parties, Alice, Bob. Each
of them performs a measurement chosen out of two: $A$($A'$) for
Alice and $B$($B'$) for Bob. The corresponding possible
outcomes are assigned as $j$ ($j'$) with $j=a,b$, respectively, and $j\in
\{-1,1\}$. Under the assumption of the existence of local hidden variables,
 the joint outcome probability after many rounds of measurements 
 is written as
\begin{equation}
P(ab|AB)=\int d\lambda~\rho(\lambda) P_{\lambda}(a|A)P_{\lambda}(b|B),
\end{equation}
where $\lambda$ is a shared local variable and $P_{\lambda}(j|J)$ is the probability that measurement $J=A,B$ 
has outcome $j=a,b$ at a set value of $\lambda$. The Bell-CHSH parameter can then be cast in the form of
\begin{equation}
\label{eq:CHSH} {\cal B}=a(b+b')+a'(b-b').
\end{equation}
It is straightforward to check that $|{\cal B}| \leq 2$. The violation of such inequality witnesses 
the failure of local hidden variable theories to describe the correlations between the outcomes of the measurements 
performed by Alice and Bob and, thus, the untenability of the assumptions of locality and realism. It is well known that if the two distributed particles 
are prepared in an entangled quantum state, the Bell-CHSH inequality can be violated by up to a factor $\sqrt2$.

This reasoning can be extended to the inclusion of a third party, say Charlie, who performs measurements $C$ and $C'$ with 
outcomes $c,c'\in\{-1,1\}$ on a third particle. In this case, it is possible to combine the outcomes of each measurements so as to build the 
parameter~\cite{merminkla}
\begin{equation}
\label{eq:Mermin}
\begin{aligned}
{\cal M}&=\frac{1}{2}{\cal B}(c+c')+\frac{1}{2}{\cal B'}(c-c')\\
&=a'bc+ab'c+abc'-a'b'c',
\end{aligned}
\end{equation}
which is bounded as $| {\cal M} | \leq 2$ according to local
hidden variable theories. Eq.~\eqref{eq:Mermin} embodies the Mermin-Klyshko (MK) 
inequality, which can be violated by quantum mechanics, thus showing tripartite quantum
nonlocality.

However, any violation of the MK inequality in Eq.~\eqref{eq:Mermin} does not
guarantee the existence of {\it genuine} tripartite nonlocality:  nonlocal correlations 
between any two parties out of three are sufficient to exceed the boundary imposed by local
hidden variable theories. To avoid this, the joint probability
for genuine tripartite nonlocality should not be reproduced by any
local hidden variable assigned on the joint measurement of any two
out of three parties. That is, we define the joint probability
\begin{equation}
\label{eq:jointp} 
P(abc|ABC)=\int d\lambda \sum{\cal P}[
\rho_{(ab)c}(\lambda) P_{\lambda}(ab|AB)P_{\lambda}(c|C)],
\end{equation}
where ${\cal P}$ performes the cyclic permutation of the triplets of indeces $(abc)$ and $(ABC)$, $P_{\lambda}(ij|IJ)$ 
stands for the joint probability of outcomes $i,j=a,b,c$ for the measurements $I,J=A,B,C$, and $\rho_{(ab)c}$ denotes the corresponding probability distribution density.
Building on this, Svetlichny proposed a combination of measurements outcomes of the form~\cite{Svetlichny}
\begin{equation}
\label{eq:Svet}
\begin{aligned}
{\cal S}&={\cal B}c \pm {\cal B'}c = {\cal M} \pm {\cal M'}\\
&=a'bc+ab'c+abc'-a'b'c'\\
&\pm a b'c'\pm a'bc' \pm a'b'c \mp abc.
\end{aligned}
\end{equation}
Again, it can be proven easily that $|{\cal S}| \leq 4$ under the assumptions behind the
Eq.~(\ref{eq:jointp}). The violation of the Svetlichny inequality \eqref{eq:Svet} thus signals genuine
tripartite nonlocality. Quantum mechanics is know to violate the bounds imposed by local hidden variables: when the probabilities entering 
Eq.~\eqref{eq:Svet} are calculated performing measurements over a tripartite GHZ 
or a W state, the Svetlichny parameter $\langle{\cal S}\rangle$ is larger than 4 (achieving the value $4\sqrt2$ over a GHZ state). 


\section{Phase space formulation via generalized quasiprobability distributions} 
\label{section:PSformalism}

In this Section we re-formulate multipartite Bell inequalities in terms of generalized
quasiprobability functions defined in the phase space of CV systems. Our approach will be based on 
the formalism introduced in Refs.~\cite{SWLee09,SWLee10} and provides the complement to the seminal analysis by Banaszek and Wodkiewicz~\cite{Banaszek99} of bipartite CV states. The $s$-parameterized quasiprobability
function for a given single-mode state $\hat{\rho}$ reads~\cite{Cahill69,SWLee10,SWLee09} 
\begin{equation}
\label{eq:sQP} 
W(\alpha; s)=\frac{2}{\pi(1-s)}\mathrm{Tr}[\hat{\rho} \hat{\Pi}(\alpha; s)],
\end{equation}
where $\hat{\Pi}(\alpha; s) = \sum^{\infty}_{n=0}[(s+1)/(s-1)]^n
\ket{\alpha,n}\bra{\alpha,n}$, and $\ket{\alpha,n}=\hat
D(\alpha)\ket{n}$ is the $n$-photon state of a boson described by annihilation and creation operator $\hat{a}$ and $\hat{a}^\dag$, displaced in phase space by the Weyl operator 
 $D(\alpha)=\exp[\alpha\hat{a}^\dag-\alpha^*\hat{a}]$. $W(\alpha; s)$ reduces to the P,
Wigner, and Husimi Q function for $s=1, 0, -1$, respectively~\cite{Moya93}.

Suppose that Alice, Bob and Charlie, independently choose one of two
observables, denoted by $\hat{A}_a$, $\hat{B}_b$, $\hat{C}_c$
respectively, where $a,b,c=1,2$, where no restriction is placed on
the number of possible outcomes. The local measurement operators
are written as
\begin{equation}
 \label{eq:LOOp}
  \hat{J}_j =\hat{O}(\delta_j; s)~\textrm{with}~\delta_j=
  \begin{cases}
 \alpha_a~&\textrm{for}~(J,j)=(A,a)\\
  \beta_b~&\textrm{for}~(J,j)=(B,b)\\
  \gamma_c~&\textrm{for}~(J,j)=(C,c)
  \end{cases}
\end{equation}
and the Hermitian operator
\begin{equation}
 \label{eq:ObOp}
  \hat{O}(\alpha;s) =
  \begin{cases}
(1-s)\hat{\Pi}(\alpha;s)+s\openone & \text{if $-1 < s \leq 0$},\\
2\hat{\Pi}(\alpha;s)-\openone & \text{if $ s \leq -1$}
\end{cases}
\end{equation}
with 
$\openone$ the identity operator. The possible measurement outcomes of $\hat{O}(\alpha;s)$
are given by the eigenvalues
\begin{eqnarray}
  \label{eq:EVal}
\lambda_n =
    \begin{cases}
(1-s)(\frac{s+1}{s-1})^n+s & \text{if $-1 < s \leq 0$},\\
2(\frac{s+1}{s-1})^n-1 & \text{if $ s \leq -1$},
\end{cases}
\end{eqnarray}
and their eigenvectors are the displaced number states. The maximum
and minimum measurement outcomes of $\hat{O}(\alpha;s)$ for any
non-positive $s$ are $\lambda_{\mathrm{max}} =-\lambda_{\mathrm{min}} = 1$. For $s=0$ we have
$\hat{O}(\alpha;0)=\hat{\Pi}(\alpha; 0)=\sum^{\infty}_{n=0}(-1)^n
\ket{\alpha,n}\bra{\alpha,n}$, which is the displaced parity operator. On the other hand, 
for $s=-1$ we have
$\hat{O}(\alpha;-1)=2\ket{\alpha}\bra{\alpha}-\openone$, {\it i.e.} a projector
onto coherent states. The MK and Svetlichny parameters can be constructed
using the measurement operators $\hat{J}_j$. Indeed, from Eq.~(\ref{eq:Mermin}), the MK parameter is now
\begin{equation}
\label{eq:Merminop} {\cal M}=
{\cal C}_{112}+{\cal C}_{121}+{\cal C}_{211}-{\cal C}_{222},
\end{equation}
where $\hat{\cal C}_{abc} =\langle\hat{A}_a\otimes\hat{B}_b\otimes\hat{C}_c\rangle$
is the correlation function for measurement outcomes. As the expectation value of any 
local observable is bounded by $1$ for any $s\le0$, the expectation value of the
MK operator given in Eq.~(\ref{eq:Merminop}) is such that $|{\cal M}| \leq 2$ for any local hidden
variable theory. From Eq.~(\ref{eq:sQP}), the expectation value of
$\hat{\Pi}(\alpha; s)$ is proportional to the $s$-parameterized
quasiprobability function as $\langle \hat{\Pi}(\alpha; s) \rangle =
\pi(1-s)W(\alpha; s)/2$. Note that we do not consider the case $s>0$
when the eigenvalues of $\hat{\Pi}(\alpha; s)$ are unbound. These considerations allow us to calculate the explicit form of the $s$-parameterized MK parameter for the two ranges of values of $s$ considered so far. Such expressions are too long and uninformative to be given here and are deferred to Appendix~{1}. 

Starting from Eq.~(\ref{eq:Svet}) and the expression in Eq~\eqref{eq:Merminps} we can easily construct the $s$-dependent Svetlichny parameter such that
\begin{equation}
\label{eq:Steve's} 
|{\cal S}_s|=|{\cal M}_s \pm {\cal M'}_s|\leq4.
\end{equation}
This is a generalization of the Svetlichny-like inequality formulated for
$s$-parameterized quasiprobability functions. For example, it
reduces to the Svetlichny inequality formulated for the Wigner
function~\cite{Li11} when $s=0$ and for the Husimi Q-function for $s=-1$. The full form of such Svetlichny 
parameters is given in Appendix~1. Using the MK and Svetlichny inequalities in
Eq.~(\ref{eq:Merminps}) and Eq.~(\ref{eq:Steve's}), we can test
genuine three mode nonlocality for arbitrary systems represented in
phase space.

In the remainder of the paper we apply our phase-space formalism to the study of a few paradigmatic examples that will serve as useful benchmarks.

\section{testing genuine nonlocality for three-mode entangled states}
\label{section:genuinenonlocality}

Here, we investigate genuine tripartite nonlocality using our phase-space formulation of Svetlichny-like inequalities by addressing three exemplary cases of three-mode entangled photon states.
Specifically, we consider the three-mode single-photon entangled state 
\begin{equation} 
\label{eq:sW}
\ket{\phi}=\sqrt{1-\frac{p}{3}}\frac{\big(\ket{0}\ket{0}\ket{1}+
\ket{0}\ket{1}\ket{0}\big)}{\sqrt2}+\sqrt{\frac{p}{3}}\ket{1}\ket{0}\ket{0},
\end{equation}
with $p\in[0,1]$, which becomes a W-state for $p=1$ and a two-mode single-photon entangled state for $p=0$. The state can be generated when a single photon enters a three-mode beam splitter (tritter). Its $s$-parameterized quasiprobability function can be written as
\begin{equation}
\label{eq:sparaSW} 
\begin{aligned}
&W_\phi(\alpha,\beta,\gamma;s)=\frac{8}{\pi^3(1-s)^3}e^{-2\frac{|\alpha|^2+|\beta|^2+|\gamma|^2}{1-s}}\\
&\times\bigg(-\frac{1+s}{1-s}+\frac{4}{(1-s)^2}\bigg|\sqrt{\frac{p}{3}}\alpha+\sqrt{\frac{1}{2}(1-\frac{p}{3})}(\beta+
\gamma)\bigg|^2\bigg).
\end{aligned}
\end{equation}
Quantum nonlocality of a two-mode single-photon state has been demonstrated theoretically~\cite{SWLee09}, but so far there have been no addressing of the genuinely nonlocal nature of its three-mode counterpart.

The second example that we consider is the three-mode squeezed vacuum (3MSV) states introduced in Ref.~\cite{PvanL01}, which can be generated by combining three single-mode squeezed states (with identical degree of squeezing $r$) at a tritter~\cite{PvanL01,Aoki03}. Its $s$-parameterized quasiprobability function is
\begin{widetext}
\begin{equation}
\label{eq:sparaSW1} 
\begin{aligned}
W_{\text{3MSV}}(\alpha,\beta,\gamma;s)=\frac{8}{\pi^3(1+s^2-2s\cosh{2r})^{3/2}}\exp\bigg[\frac{2}{3(1+s^2-2s\cosh{2r})}\bigg\{3(s-\cosh{2r})(|\alpha|^2+|\beta|^2+|\gamma|^2)\\
+(\alpha_i^2+\beta_i^2+\gamma_i^2-4\alpha_i\beta_i-4\beta_i\gamma_i-4\gamma_i\alpha_i-\alpha_r^2-\beta_r^2-\gamma_r^2+4\alpha_r\beta_r+4\beta_r\gamma_r+4\gamma_r\alpha_r)
\sinh{2r}\bigg\}\bigg],
\end{aligned}
\end{equation}
\end{widetext}
where $\delta_r$ and $\delta_i$ denote the real and imaginary parts of $\delta=\alpha,\beta,\gamma$. Such Gaussian state has been introduced as the CV version of the GHZ entangled state~\cite{PvanL01}, although for small squeezing it contains some residual two-mode entanglement and it can thus be also regarded as a representative of W-class entanglement~\cite{PvanL02}.

Finally, we address entangled coherent states (ECS) of the form
\begin{equation}
\label{eq:cohGHZ} 
\ket{\text{ECS}}={\cal N}(\ket{\zeta}\ket{\zeta}\ket{\zeta}-\ket{-\zeta}\ket{-\zeta}\ket{-\zeta}),
\end{equation}
where $\zeta$ is the amplitude of a single coherent state (we assume for simplicity that $\zeta\in{\mathbb R}$) and $\cal N$ is a normalization constant. A scheme for the generation of these states was introduced in Ref.~\cite{Jeong06}. Although states $\ket{\text{ECS}}$ have been regarded as the entangled coherent state version of GHZ-class entanglement~\cite{Jeong06}, it has been also known that, in the limit of $\zeta \to 0$, Eq.~\eqref{eq:cohGHZ} reduces to Eq.~\eqref{eq:sparaSW} with $p=1$. Therefore, depending on the amplitude of the state components, it might be possible to identify two distinct behaviors of the Svetlichny parameter, associated with  either GHZ or W-class entanglement. 
The $s$-parameterized quasiprobability function of $\ket{\text{ECS}}$ is 
\begin{equation}
\label{eq:sparaSW2}
\begin{aligned}
&W_{\text{ecs}}(\alpha,\beta,\gamma;s){=}\frac{8{\cal
N}^2}{\pi^3(1-s)^3}\left(\sum_{g=\pm}e^{-2\frac{|\alpha+g\zeta|^2+|\beta+g\zeta|^2+|\gamma+g\zeta|^2}{1-s}}\right.\\
&
\left.-2e^{2\frac{3s\zeta^3-|\alpha|^2-|\beta|^2-|\gamma|^2}{1-s}}\cos\bigg[\frac{4\zeta(\alpha_i+\beta_i+\gamma_i)}{1-s}\bigg]\right).
\end{aligned}
\end{equation}
It would also be tempting to study  the ECS states ${\cal
N'}(\ket{0}\ket{0}\ket{\zeta}+\ket{0}\ket{\zeta}\ket{0}+\ket{\zeta}\ket{0}\ket{0})$, sometimes referred to as W-type ECS~\cite{Jeong06}. However, the non locality of this class of states cannot be revealed by relying on Bell-like tests based on local displacement in phase space, as considered here, and require the use of local operations based on Kerr-like nonlinearities instead~\cite{Jeong06}.

The three paradigmatic classes of states introduced above will be now studied against their tripartite nonlocal character. Later on in this Section we consider the behaviour of both MK and Svetlichny-like parameters against the influences of dissipation affecting such states, which will allow us to comment on the robustness of genuine tripartite nonlocality. 

\begin{figure}[b!]
\includegraphics[width=\linewidth]{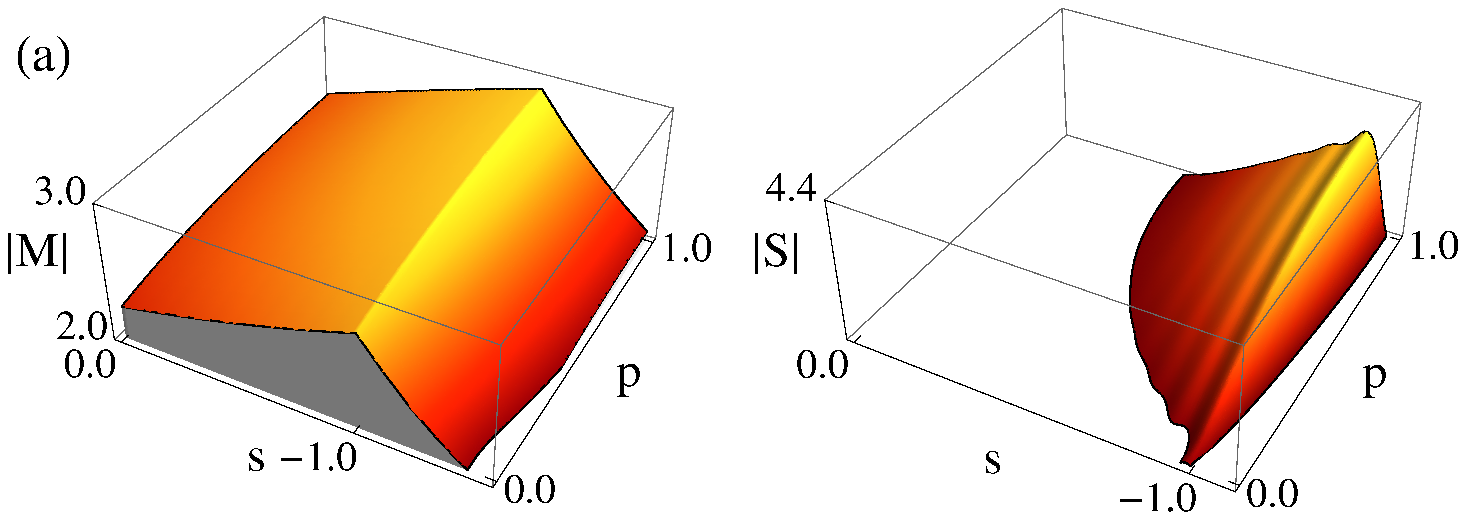}
\includegraphics[width=\linewidth]{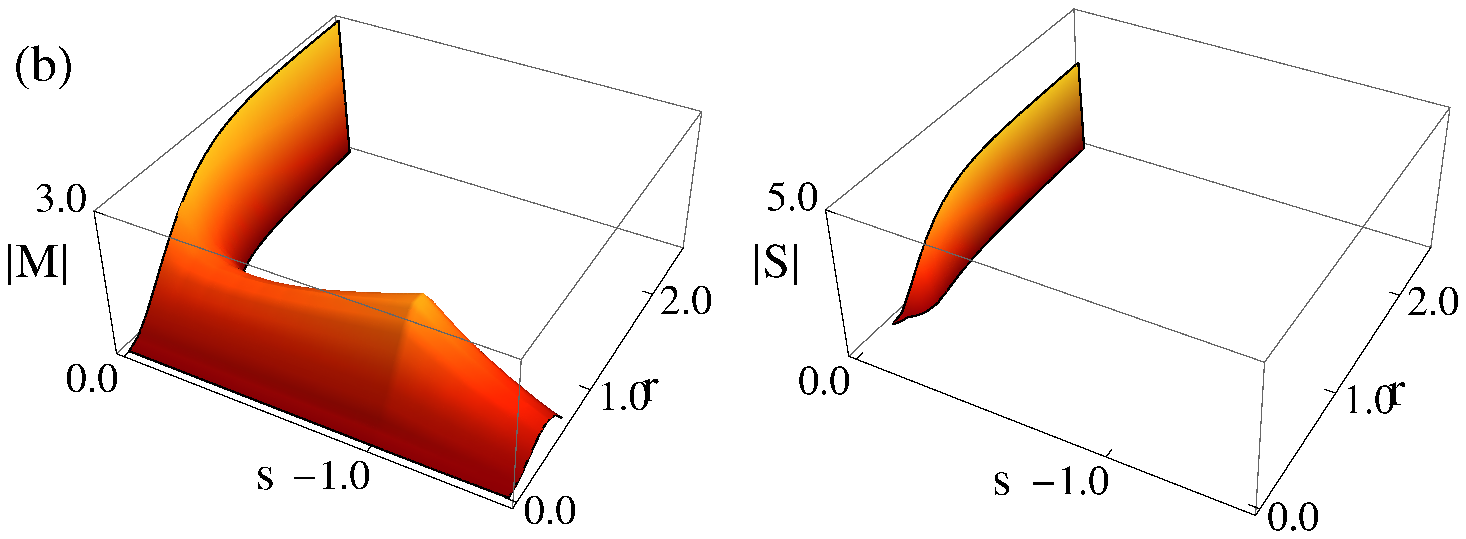}
\includegraphics[width=\linewidth]{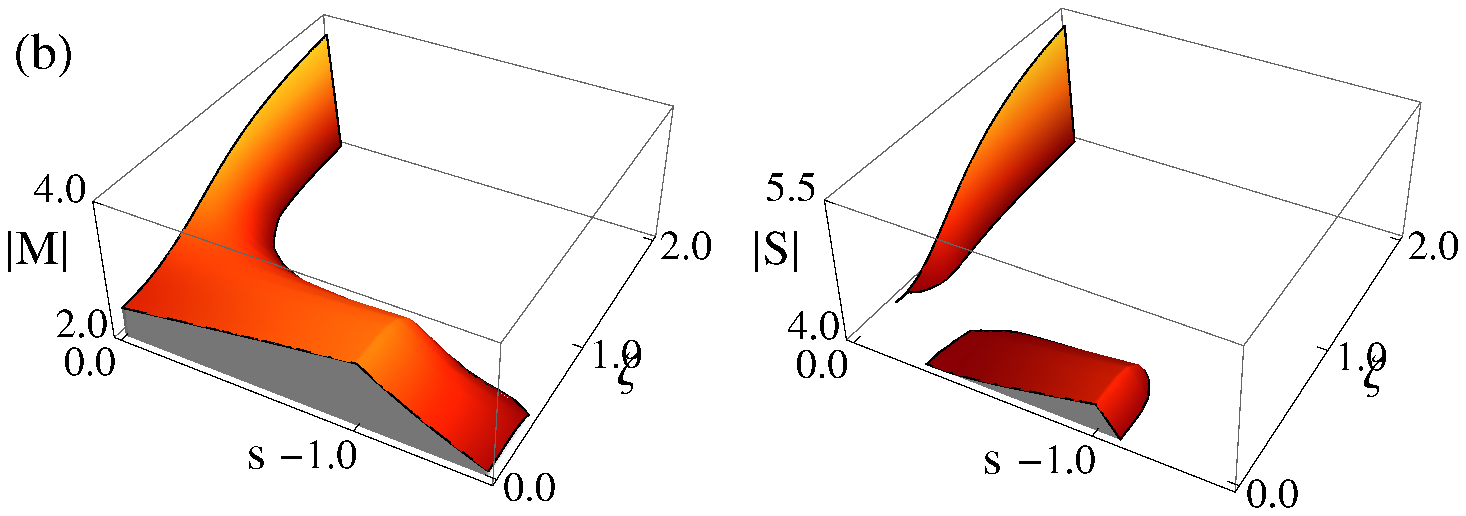}
\caption{Behavior of the $s$-parameterized MK and Svetlichny parameter $|{\cal M}|$ and $|{\cal S}|$ for (a) the single-photon W state $\ket{\phi}$ (with variable $p$), (b) the three-mode squeezed vacuum state ($r$ being the degree of squeezing), and (c) the GHZ-type ECS ($\zeta$ is the amplitude of each coherent-state component).}\label{fig1}
\end{figure}

\subsection{Demonstrating genuine three-mode nonlocality in phase space}

We start investigating the violations of both MK and Svetlichny type inequalities for each of the entangled states discussed above. As in our formalism quantum nonlocality is independent of the local measurement setting associated with parameter $s$, the violation of a Svetlichny-like inequality for a suitably chosen $s$ guarantees the presence of genuine three-mode nonlocality in the specific entangled state being studied. In the following, we refer to the genuine tripartite nonlocality exhibited by states belonging to the GHZ- and W-class entangled states as {\em GHZ-type} and {\em W-type nonlocality}, respectively.

The genuine tripartite W-type nonlocal nature of a single-photon entangled state $\ket{\phi}$ for proper choices of $p$ can be appreciated from the right panel of Fig.~\ref{fig1}(a): Both the MK and Svetlichny inequalities are violated maximally by a test designed using the Husimi Q function ({\em i.e.} for $s=-1$). However, while the MK inequality is violated for all values of $s$ (Fig.~\ref{fig1} (a), left panel), the Svetlichny one is violated only close to $s=-1$, implying the relevance of the choice of suitable local measurements to demonstrate genuine three-mode nonlocality. Clearly, when $p=0$ the state is solely endowed with two-mode entanglement, which yields the violation of the MK test only. The Svetlichny parameter depends monotonically on $p$ and  achieves its maximum value at $p=1$.

For the three-mode squeezed vacuum states, the Svetlichny inequality can be violated only in a region of values very close to $s=0$ [cf. Fig.~\ref{fig1}(b), right panel], where the dependence of ${\cal S}$ on the degree of squeezing is monotonic and where observables embodied by local displaced parity operations (as it is the case for phase-space formulation of Bell-like tests based on the Wigner function) are optimal. From the observation in Ref.~\cite{PvanL02} that the entanglement of any two-mode subsystem disappears for sufficiently large degree of squeezing $r$, we claim for GHZ-type nonlocality, in this case. 
For small values of $r$, though, W-class entanglement becomes dominant in the state. For example, it is known that a three-mode squeezed vacuum state with $r \simeq0.5$ contains W-type entanglement dominantly as its residual two-mode entanglement becomes maximum ~\cite{PvanL02}. This is in agreement with our analysis of the MK parameter ${\cal M}$, whose formulation in terms of the Q function ($s =-1$) maximizes the degree of violation of the MK inequality, as shown in the left panel of Fig.~\ref{fig1}(b). However, note that W-type nonlocality does not appear in this range of $r$, implying that not all W-class entanglement can yield genuine W-type nonlocality in phase space.

\begin{figure}[b!]
\includegraphics[width=.85\linewidth]{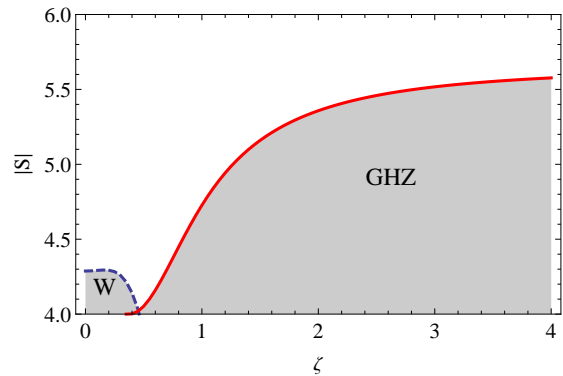}
\caption{Genuine tripartite nonlocality phase diagram for a three-mode ECS. We plot the Svetlichny parameter $|{\cal S}|$ against the amplitude of the coherent-state components $\zeta$. The W-type and GHZ-type nonlocal characters, best revealed by tests built on the Husimi ($s=-1$, dashed line) and Wigner function ($s=0$, solid line), respectively, cross for $\zeta\simeq0.455$.}
\label{fig1bis}
\end{figure}

For the ECS in Eq.~(\ref{eq:cohGHZ}), both GHZ and W-type nonlocality can emerge with an appropriate choice of $s$, as shown in Fig.~\ref{fig1}(c). For small amplitudes $\zeta$, genuine tripartite nonlocality is observed for a test built using the Q function. This part of the tripartite non locality phase diagram for an ECS is originated by the W-class entanglement of a single-photon state and thus reveals W-type nonlocality. Needless to say, the results associated with $\zeta \rightarrow 0$ are the same as those obtained for $p=1$ in state $\ket{\phi}$ and shown in Fig.~\ref{fig1}(a) (right panel). On the other hand, GHZ-type nonlocality is achieved in a very narrow region close to $s=0$. In this case, a test based on the Wigner function ({\it i.e.} the use of local parity measurements) is optimal for demonstrating CV GHZ-type nonlocality. As shown in Fig.~\ref{fig1bis}, the W- and GHZ-type nonlocal characters are interchanged at $\zeta\simeq0.455$; past this point, the W type (best revealed by taking $s=-1$, {\it i.e.} using a test based on the Husimi Q function) disappears to leave room for a GHZ-type character (to be tested by the Wigner function). Therefore, genuine tripartite nonlocality can be observed for all ranges of $\zeta$ with suitably chosen local measurement setting.

\section{Effects of detection inefficiency and local damping on genuine nonlocality}
\label{section:noiseffcts}

Here we investigate the effect of the two important detrimental mechanisms on the genuine three-mode nonlocality demonstrated in Sec.~\ref{section:genuinenonlocality}. Realistic implementations of CV nonlocality tests should take into account both the inefficiency of the detectors used to collect the statistics and the potential interaction of each local mode participating to the state being probed with environmental baths. Both effects can be effectively described as changes of the parameter $s$ used in our formalism. In particular, as it is shown in Appendix~2, the single-mode $s$-parameterized quasiprobability function measured by a detector with efficiency $\eta\in[0,1]$ is given by a $1/\eta$ rescaling of the {\it ideal} quasiprobability function that should then be parameterized by~\cite{Banaszek96}
\begin{eqnarray}
\label{eq:efficiencys} s'=-\frac{1-s-\eta}{\eta}.
\end{eqnarray}
Therefore, the MK- and Svetlichny-type inequalities under the effect of detection noise can be formulated by replacing the quasiprobability distribution in Eq.~(\ref{eq:Merminps}) and Eq.~(\ref{eq:Steve's}) with the {\it measured} quasiprobability functions, {\em e.g.} $W_3(\alpha,\beta,\gamma;s)\rightarrow W_3(\alpha,\beta,\gamma;s')/(\eta_a \eta_b \eta_c)~(j=1,2,3)$ with $\eta_{a,b,c}$ the detection efficiency of the detectors at Alice's, Bob's, and Charlie's site, respectively (the marginals being calculated accordingly). For simplicity, in what follows we will restrict the attention to the case of $\eta_{a,b,c}=\eta$. Similar transformations are enough to take into account the effects of the interactions of each local mode with its own thermal reservoir (at equilibrium with $\bar{n}$ thermal photons) through non-zero temperature amplitude-damping mechanisms~\cite{NielsenChuang} occurring at rate $\Gamma$ (assumed to be the same for all modes, for the sake of argument). In this scale, we should operate the replacement
\begin{equation}
\label{transf}
W_3(\alpha,\beta,\gamma;s)\to \frac{1}{t^6(\tau)}W_3\left(\frac{\alpha}{t(\tau)},\frac{\beta}{t(\tau)},\frac{\gamma}{t(\tau)};s'(\tau)\right)
\end{equation}
with $t(\tau)=\sqrt{1-r^2(\tau)}=\sqrt{e^{-\Gamma\tau}}$, $\tau$ the mode-bath interaction time and
$$s'(\tau)=\frac{s-r^2(\tau)(1+2\bar{n})}{t^2(\tau)}.$$
Equipped with this tools, we can now address the effects of both detection inefficiency and local damping on the W- and GHZ-type nonlocality.

\begin{figure}
\begin{center}
\includegraphics[width=0.8\linewidth]{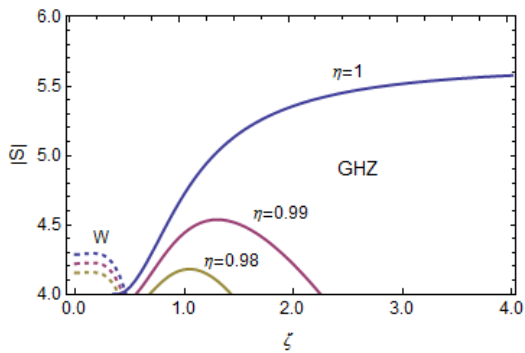}
\caption{Tripartite non locality phase diagram for a three-mode ECS against the detection efficiency $\eta$. As in Fig.~\ref{fig1bis}, dashed (solid) traits are for $s=-1$ ($s=0$)}\label{fig2}
\end{center}
\end{figure}

\begin{figure}[b]
\begin{center}
\includegraphics[width=1\linewidth]{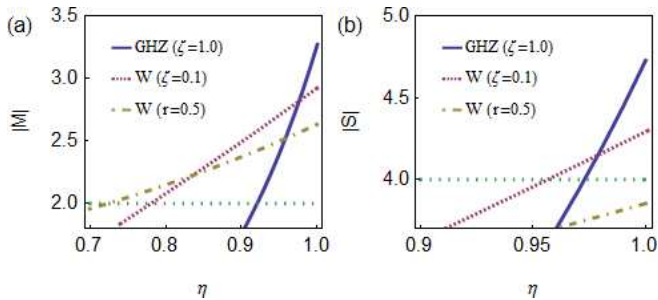}
\caption{Behavior of the MK parameter [panel (a)] and the Svetlichny one[panel (b)] against the section efficiency $\eta$ for a three-mode ECS ($\zeta=1.0, 0.1$) and a three-mode squeezed state ($r=0.5$).}\label{fig3}
\end{center}
\end{figure}

\begin{figure}
\begin{center}
\includegraphics[width=0.9\linewidth]{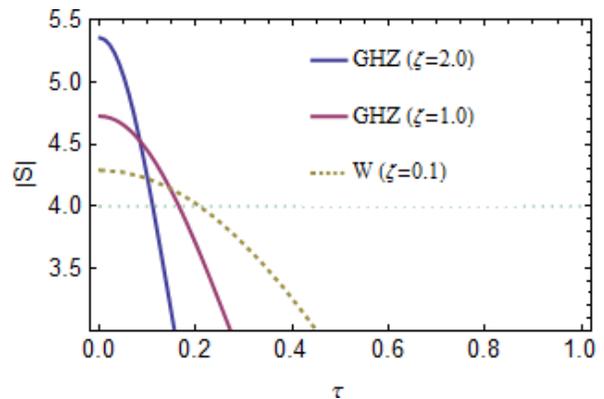}
\caption{Dynamics of genuine GHZ-type ($\zeta=1,2$) and W-type ($\zeta=0.1$) nonlocality of a three-mode ECS interacting with individual thermal baths at equilibrium (mean occupation  number $\bar{n}=0$).}\label{fig4}
\end{center}
\end{figure}

In Fig.~\ref{fig2}, we plot the effect induced by imperfect detection on the behavior of the Svetlichny parameter for a three-mode ECS. Linking to our previous considerations, we focus on the trade-off between W- and GHZ-type nonlocality. We use the trends shown in Fig.~\ref{fig1bis} as a benchmark for the functional behaviour associated with $\eta<1$. Clearly, GHZ-type nonlocality is somehow more prone to the effects of imperfect detection, as the solid-line traits in Fig.~\ref{fig2} (associated with Wigner function-based non-locality tests, {\it i.e.} for $s=0$) disappear quickly for even very weakly inefficient detectors. The W-type nonlocal character appears to be quite more robust: $\eta \gtrsim 0.97$ is required to demonstrate GHZ-type nonlocality for the entangled coherent state with $\zeta=1.0$, while W-type nonlocality for the state with $\zeta=0.1$ can be observed with $\eta \gtrsim 0.955$. Needless to say, this fragility is partly due to the large amplitude of the coherent-state components needed to test GHZ-type nonlocality. Continuing our analysis, we see that, expectedly, in comparison with the MK parameter, the Svetlichny one is quite more sensitive. In Fig.~\ref{fig3}, we find the threshold in the detection efficiency that is needed in order to observe the violation of both MK and Svetlichny inequalities by a three-mode ECS. The verification of genuine W-type nonlocality with a W-class ECS  state ($\zeta=0.1$) requires $\eta \gtrsim 0.955$, while $\eta \gtrsim 0.78$ is enough to observe the violation of the MK inequality. These considerations hold, qualitatively, also when the effector local damping is considered [cf. Fig.~\ref{fig4}]

Our final confederation in this respect is that the robustness shown by the MK parameter for a given three-mode state does not always imply an equally robust Svetlichny parameter. For example,  the three-mode squeezed vacuum state with degree of squeezing $r=0.5$ exhibits maximal W-class entanglement~\cite{PvanL02} and has an associated MK parameter that is more resilient to detection inefficiencies than the ECS with amplitude $\zeta=0.1$ as shown in Fig.~\ref{fig3}. However, it does not yield any genuine W-type nonlocality, even with perfect detectors, in contrast with the some how more fragile ECS.

\section{Physical settings}
\label{section:Settings}

The progresses made in the last ten years in the design, manipulation and control of mesoscopic quantum systems consisting of hybrid components has now made possible the experimental realisation of interfaces between devises as diverse as ultra cold atoms and high-finesse cavities~\cite{Esslingerreview} or micro mechanical oscillators~\cite{hunger}. On the other hand, the development of well-acquired techniques for the spacial confinement of charge atoms has now reached outstanding levels allowing for the quantum processing of tens of individual particles~\cite{Blatt}. This is very interesting for the purposes of our study. 

Indeed, although throughout this paper we have used the language typical of all-optical implementation of CV systems, the formal apparatus built in our study applies, needless to say, to any effective bosonic system, regardless of its physical embodiment. This makes the state of ensembles of cold and/or ultra cold atoms, mico/nanomechanical oscillators and the vibrational degree of freedom of trapped particles, as well as the interface between such systems, perfectly suited to be tested via the phase-space approach described here. In fact, our formalism holds the potential to embody a very powerful instrument for the analysis of the sharing of quantum correlations among the elements of a multipartite hybrid system comprising effective bosonic modes of various natures, as well as truly many-body arrays of identical trapped particles whose vibrational degrees of freedom would embody the register of bosons to study. 

In particular, we can explicitly mention linear~\cite{Blatt}, as well as planar~\cite{schmidtkaler} and multipole ion traps~\cite{gerlich}, which are able to accommodate ions trapped in unidimensional, as well as bidimensional configurations (so-called {\it ionic crystals}: bidimensional structures are formed either via crystallization of explicit confinement of multi-ion arrays). In this case, the vibrational degrees of freedom of the trapped ions, which are effectively interacting  due to Coulomb repulsion, give rise to interesting long-range couplings that are likely to result in multipartite states whose nonlocal character can well be studied by the means of our tools~\cite{Li11, Serafini}. Similar considerations hold for ultra cold atoms loading an intracavity double-well potential, where in the so-called two-mode approximation~\cite{Milburn}, a tripartite bosonic state of a cavity field and two atomic modes can be established. 

\begin{figure}
\begin{center}
\includegraphics[width=\linewidth]{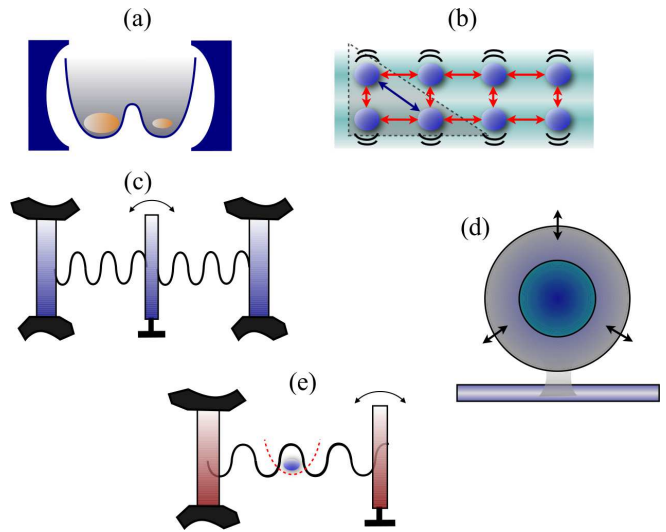}
\caption{Exemplary physical systems where our phase-space formalism could be applied to investigate genuine multipartite nonlocality: (a) An ultra cold atomic system loaded into an intra-cavity double-well potential~\cite{Margherita}; (b) a Coulomb crystal of ions confined on a planar on-chip trap (the dashed triangle identified a subsystem of three ions whose state can be studied with our proposed tools. Red [Blue] arrows show nearest-neighbor [next-to-nearest-neighbor] interactions)~\cite{gerlich}; (c) a mechanical membrane in the middle of a cavity~\cite{Paternostro2007}; (d) a doped micro-toroid coupled to a fibre~\cite{Xuereb2012}; (e) a hybrid optomechanical device including a mechanical mode and a trapped particle~\cite{Isart} or Bose-Einstein condensate~\cite{maurogabriele}.}
\label{fig5}
\end{center}
\end{figure}

Finally, it will be interesting to address the case of multipartite optomechanical devices comprising a mechanical system interacting with multiple optical modes, and hybrid optomechanical systems comprising vibrating mechanical modes, such as the situations addressed explicitly in Refs.~\cite{Paternostro2007,Xuereb2012}, as well as hybrid configurations including mechanical systems that are coupled to ultracold atoms (such as a Bose-Einstein condensate)~\cite{hunger,maurogabriele,maurogabriele1} or an optically trapped dielectric sphere~\cite{Isart}. In Fig.~\ref{fig5} we show the sketch of some of such instances of physical systems, which are discussed in details in the references mentioned in the figure caption, to which we refer. 

\section{Conclusions}
\label{section:discon}

We have formulated Svetlichny-type Bell tests for $s$-parameterized quasiprobability functions in phase space and tested them using interesting three-mode CV states. GHZ- and W-type nonlocality can be distinguished, with our tools, by properly adjusting the value of $s$. GHZ-type nonlocality with many photons appears to be witnessed using local parity measurements, while single-photon W-type nonlocality requires on-off measurements ({\it i.e.} the Husimi function). This reflects the fact that testing genuine multipartite nonlocality would require the pondered choice of local measurement settings, depending on the type of state being studied, in contrast to the violation of MK, which occurs for a wide ranges of values of $s$. 

We observe that not all W-class entanglement can violate a Svetlichny inequality, indicating that the nonlocal character of genuinely multi-mode entangled states does not always coincide with genuine multi-mode nonlocality. 
In studying the behaviour of Svetlichny parameters for tripartite entangled states, we found that ECSs are useful to demonstrate genuine three-mode nonlocality in the whole range of amplitudes of the state components and that, the stronger GHZ-type nonlocality, the more fragile it is under detrimental effects. Differently, a pronounced W-type nonlocal character is robust to detection inefficiency and local damping.

Our results consistently show that multi-photon states  are endowed with very fragile genuinely multipartite nonlocality content. As our analysis reveals, W-type nonlocality is exhibited by states of a small amplitude (asymptotically, multimode single-photon states), while GHZ-type nonlocality is inherent in GHZ-class entangled states of many photons. This explains why the GHZ-type nonlocality is found to be more prone to environmental actions.

Our phase-space inequalities embody useful and powerful tools for the investigation of multipartite quantumness in quantum technology settings of current experimental interests. As discussed in this paper, there is large fan of physical configurations, involving effective bosons of various physical embodiments, that can benefit from the application of phase-space methods for the tests of multipartite nonlocality,  a task that we will pursue in the close future. 



\acknowledgments
MP is grateful to Th. Busch, C. Cormick, J. Goold, and J. Li for discussions on topics related to the subject of this work. This work was supported by the National Research Foundation of Korea (NRF) grant funded by the Korean Government (No.
3348-20100018), the World Class University program, and the UK EPSRC through a Career Acceleration Fellowship and the ``New Directions for EPSRC Research Leaders" initiative (EP/G004759/1).

\renewcommand{\theequation}{A-\arabic{equation}}
\setcounter{equation}{0}  
\section*{Appendix 1}
\label{A1}

In this Appendix we report the explicit form of the $s$-parameterized MK parameters for the two ranges of values of $s$ stated in the main body of the paper. Following the approach described in Sec.~\ref{section:PSformalism}, we obtain the MK parameters 
\begin{widetext}
\begin{equation}
\label{eq:Merminps}
\begin{aligned}
  {\cal M}_{\{-1<s\leq0\}}&=
  \frac{\pi^3(1-s)^6}{8}{\cal D}_{W_3}({\bm \alpha},{\bm \beta},{\bm\gamma};s)+\frac{\pi^2(1-s)^4s}{4}[{\cal D}_{W_2}({\bm \alpha},{\bm \beta};s)+{\cal D}_{W_2}({\bm \beta},{\bm\gamma};s)+{\cal D}_{W_2}({\bm \gamma},{\bm\alpha};s)]\\
&+\pi(1-s)^2s^2\sum_{\delta=\alpha,\beta,\gamma}{W_1}(\delta;s)+2s^3,\\\\
  {\cal M}_{\{s\leq-1\}}&=
  \pi^3(1-s)^3{\cal D}_{W_3}({\bm \alpha},{\bm \beta},{\bm\gamma};s)+\pi^2(1-s)^2[{\cal D}_{W_2}({\bm \alpha},{\bm \beta};s)+{\cal D}_{W_2}({\bm \beta},{\bm\gamma};s)+{\cal D}_{W_3}({\bm \gamma},{\bm \alpha};s)]\\
&+2\pi(1-s)\sum_{\delta=\alpha,\beta,\gamma}{W_1}(\delta;s)-2,
\end{aligned}
\end{equation}
\end{widetext}
where we have introduced the function
\begin{equation}
\begin{aligned}
{\cal D}_{W_3}({\bm \alpha},{\bm \beta},{\bm\gamma};s)&={W_3}(\alpha,\beta,\gamma';s)+{W_3}(\alpha,\beta',\gamma;s)\\
&+{W_3}(\alpha',\beta,\gamma;s)-{W_3}(\alpha',\beta',\gamma';s)
\end{aligned}
\end{equation}
and, say
\begin{equation}
\begin{aligned}
{\cal D}_{W_2}({\bm \alpha},{\bm \beta};s)&={W_2}(\alpha,\beta;s)+{W_2}(\alpha,\beta';s)\\
&+{W_2}(\alpha',\beta;s)-{W_2}(\alpha',\beta';s).
\end{aligned}
\end{equation}
which are defined in terms of the set of measurement outcomes ${\bm\delta}=\{\delta,\delta'\}$ with $\delta=\alpha,\beta,\gamma$, 
the three-mode correlator
$$W_3(\alpha,\beta,\gamma; s)=\frac{8}{\pi^3(1-s)^3}\mathrm{Tr}[\hat{\rho} \hat{\Pi}(\alpha; s)\otimes\hat{\Pi}(\beta; s)\otimes\hat{\Pi}(\gamma; s)]$$ 
and its marginals 
\begin{equation}
\begin{aligned}
\nonumber
&W_2(\alpha,\beta;s)=\int  W_3(\alpha,\beta,\gamma;s)d^2\gamma,\\
&W_1(\alpha;s)=\int W_3(\alpha,\beta,\gamma;s)d^2\beta d^2\gamma.
\end{aligned}
\end{equation}

Finally, we provide the form of the Svetlichny parameter expressed in terms of the Wigner and Husimi Q functions, which read
\begin{equation}
\label{eq:svetpsW} 
{\cal S}_{s=0}=\frac{\pi^3}{8}[{\cal D}_{W_3}({\bm\alpha},{\bm\beta},{\bm\gamma};0)\pm{\cal D}'_{W_3}({\bm\alpha},{\bm\beta},{\bm\gamma};0)],
\end{equation}
\begin{equation}
\begin{aligned}
{\cal S}_{s=-1}&=8\pi^3[{\cal D}_{Q_3}({\bm \alpha},{\bm \beta},{\bm \gamma})+{\cal D}'_{Q_3}({\bm \alpha},{\bm \beta},{\bm \gamma})]\\
&+8\pi^2[Q(\alpha,\beta')+Q(\beta,\gamma')+Q(\gamma,\alpha')]\\
&+8\pi^2[Q(\alpha',\beta)+Q(\beta',\gamma)+Q(\gamma',\alpha)]\\
&+4\pi\sum_{\delta=\alpha,\beta,\gamma}[Q(\delta)+Q(\delta')],
\end{aligned}
\end{equation}
respectively, where 
\begin{equation}
\begin{aligned}
{\cal D}'_{W_3}({\bm \alpha},{\bm \beta},{\bm\gamma};0)&=W(\alpha',\beta',\gamma)+
W(\alpha',\beta,\gamma')\\
&+ W(\alpha,\beta',\gamma')-W(\alpha,\beta,\gamma).
\end{aligned}
\end{equation}

\renewcommand{\theequation}{B-\arabic{equation}}
\setcounter{equation}{0}  
\section*{Appendix 2}
\label{B1}

In this Appendix we provide the formal details behind the replacements to be operated on the $s$-parameterized quasi probability distributions under the effects of inefficient detectors and local amplitude damping at non-zero temperature. 


We start considering the effects of imperfect detection, which can be in general described using a virtual beam splitter placed in front of a detector with perfect efficiency. Detection efficiency is thus associated with the transmittivity $\eta$ of the virtual beam splitter, which changes the true photon-number distribution $P(n)$ of a signal into the measured quantity  
$P_{\eta}(m)=\sum^{\infty}_{n=m}P(n)\binom{n}{m}(1-\eta)^{n-m}\eta^m$ \cite{pndist}. Correspondingly, the measured quasiprobability function at the origin of phase-space ({\it i.e.} for $\alpha=0$) reads
\begin{equation}
 \label{eq:Wexp}
 \begin{aligned}
  W_{\eta}(0;s)&=\frac{2}{\pi(1-s)}\sum^{\infty}_{m=0}\left(\frac{s+1}{s-1}\right)^mP_{\eta}(m)\\
  &=\frac{W\left(0;-\frac{1-s-\eta}{\eta} \right)}{\eta}.
\end{aligned}
\end{equation}
The $s$-parameterized quasiprobability function measured by a detector with efficiency $\eta$ can thus be identified with a rescaled quasiprobability function characterised by the inefficiency-dependent parameter \cite{Banaszek96}
\begin{equation}
\label{eq:efficiencys1} s'=-\frac{1-s-\eta}{\eta}.
\end{equation}

Let us now describe the state evolution in local thermal environments. The scope is to account for such dynamics as the changes of $s$. The effect of the local thermal baths can be  modelled by the mixture, at a beam splitter, of the state of the mode under scrutiny with a thermal field. 
This help modelling the dynamics of the three modes in terms of a Fokker-Planck equation for the $s$-parameterized quasiprobability distribution function reading~\cite{Jeong2000},
\begin{equation}
\label{eq:FPeq} 
\begin{aligned}
&\frac{\partial
W(\alpha,\beta,\gamma;s;\tau)}{\partial
\tau}=\frac{\Gamma}{2}\sum_{\delta=\alpha,\beta,\gamma}\bigg[
\frac{\partial}{\partial \delta}\delta+\frac{\partial}{\partial
\delta^*}\delta^*\bigg]W(\alpha,\beta,\gamma;s;\tau)\\
&+{\Gamma}\left(\frac{1}{2}+\bar{n}\right)
\sum_{\delta=\alpha,\beta,\gamma}
\frac{\partial^2}{\partial
\delta \partial \delta^*}W(\alpha,\beta,\gamma;s;\tau),
\end{aligned}
\end{equation}
where $\tau$ is the system-bath interaction time and $\Gamma$ is the energy decay rate into the environmental baths each of average thermal photon number $\bar{n}$. The evolution can be cast into the form of a convolution between the quasi probability function of the three system modes and those of the three environmental baths at thermal equilibrium. That is
\begin{equation}
\label{eq:solFPeq}
\begin{aligned}
&W(\alpha,\beta,\gamma;s;\tau)=\frac{1}{t^6(\tau)}\int d^2 \alpha'
d^2 \beta' d^2 \gamma'
\Pi_{\delta'}W^{\mathrm{th}}(\delta';s)\\
&\times W\bigg(\frac{\alpha-r(\tau)\alpha'}{t(\tau)},
\frac{\beta-r(\tau)\beta'}{t(\tau)},\frac{\gamma-r(\tau)\gamma'}{t(\tau)};s;0
\bigg),
\end{aligned}
\end{equation}
where $r(\tau)$ and $t(\tau)$ have been introduced in the body of the paper, and
\begin{equation}
W^{\mathrm{th}}(\alpha;s)=\frac{2}{\pi(2\bar{n}+1-s)}e^{-\frac{2|\beta|^2}{2\bar{n}+1-s}}
\end{equation}
is the $s$-parameterized quasiprobability function for the thermal state of average thermal photon number $\bar{n}$. The environmental effects can thus be identified with temporal changes of the quasiprobability function as stated in Eq.~\eqref{transf}, which concludes our formal assessment. 

\end{document}